\documentclass[letter,12pt]{article}
\pdfoutput=1 

\usepackage{jheppub} 

\usepackage[T1]{fontenc} 

\usepackage{gensymb}

\usepackage{caption}
\usepackage{subcaption}
\usepackage{graphicx}

\usepackage{tikz}
\usetikzlibrary{decorations.pathmorphing}
\usetikzlibrary{positioning,decorations.pathreplacing,shapes}

\usetikzlibrary{calc}

\usepackage{verbatim}
\tikzset{snake it/.style={decorate, decoration=snake}}

\newcommand{\tikzAngleOfLine}{\tikz@AngleOfLine}
  \def\tikz@AngleOfLine(#1)(#2)#3{%
  \pgfmathanglebetweenpoints{%
    \pgfpointanchor{#1}{center}}{%
    \pgfpointanchor{#2}{center}}
  \pgfmathsetmacro{#3}{\pgfmathresult}%
  }

\def\bea{\begin{eqnarray}}
\def\eea{\end{eqnarray}}

\title{\Huge Rate of cluster decomposition via Fermat-Steiner point}

\author[a,b]{Alexander Avdoshkin,}
\author[b,c]{Lev Astrakhantsev,}
\author[d,e]{Anatoly Dymarsky,}
\author[f]{and Michael Smolkin}

\affiliation[a]{Department of Physics, University of California, Berkeley, CA, USA 94720\\}
\affiliation[b]{Institute for Theoretical and Experimental Physics,\\B. Cheremushkinskaya, 25, 117218, Moscow, Russia\\}
\affiliation[c]{Moscow Institute of Physics and Technology,\\Institutskii per, 9, 141700, Dolgoprudny, Russia\\}
\affiliation[d]{University of Kentucky,\\Lexington, KY, USA 40506\\}
\affiliation[e]{Skolkovo Institute of Science and Technology,\\Skolkovo Innovation Center, Moscow, Russia\\}
\affiliation[f]{The Racah Institute of Physics, The Hebrew University of Jerusalem, \\ Jerusalem 91904, Israel}

\emailAdd{alexander\_avdoshkin@berkeley.edu}
\emailAdd{levastrahancev@mail.ru}
\emailAdd{a.dymarsky@uky.edu}
\emailAdd{michael.smolkin@mail.huji.ac.il}

\abstract{In quantum field theory with a mass gap correlation function between two spatially separated operators decays exponentially with the distance. This fundamental result immediately implies an exponential suppression of all higher point correlation functions, but the predicted exponent is not optimal. We argue that in a general quantum field theory the optimal suppression of a three-point function is  determined by total distance from the operator locations to the Fermat-Steiner point. Similarly, for the higher point functions we conjecture the optimal exponent is determined by the solution of the Euclidean Steiner tree problem. We discuss how our results constrain operator spreading in relativistic theories.}

\begin{document} 
\maketitle
\flushbottom

\section{Introduction}
\label{sec:intro}
\begin{figure}[t]
\centering
\begin{tikzpicture}
\draw
    (0,0) coordinate (a) node[below] {$\vec{x}_3=(0,0)$}
    -- (9,0) coordinate (b) node[below] {$\vec{x}_2=(\ell_1,0)$}
    -- (6,3) coordinate (c) node[above] {$\vec{x}_1=(a,b)$}
    -- (a);
\draw (4.5,-0.3) coordinate () node[below] {$\ell_1$};
\draw (7.7,1.9) coordinate () node {$\ell_3$};
\draw (3,1.9) coordinate () node[left] {$\ell_2$};
\node at (0,0) {\textbullet};
\node at (6,2.98) {\textbullet};
\node at (9,0) {\textbullet};
\end{tikzpicture}
\caption{When the triangle inequality is satisfied, $\ell_2+\ell_3>\ell_1$, the points belong to a spatial plane and all three times can be chosen to be zero. By choosing an appropriate reference frame the points can be brought to the  configuration  shown in this picture.}
\label{fig:triangle}
\end{figure}
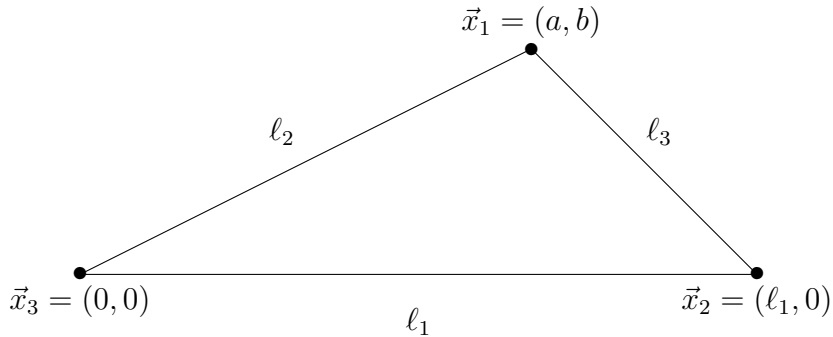

Cluster decomposition of vacuum correlation functions is one of the basic results of quantum field theory, which underlines the locality of interactions \cite{streater2016pct}. When a theory has a mass gap $m$, connected correlations between spatially separated operators decay exponentially with the distance,
\bea
\label{2pt}
\langle \phi(x) \phi(0)\rangle  \sim e^{-m |x|}, \qquad {\rm for}\ \ \ m|x|\gg 1.
\eea
In a relativistic case this fundamental result readily follows from i.e.~K\"all\'en-Lehmann spectral representation and can be established in a number of ways \cite{weinberg1995quantum}. For the discrete lattice systems a similar result applies, but the derivation is much more involved \cite{hastings2004lieb,hastings2010locality}.
From here it immediately follows that the connected correlator of several spatially separated operators is also exponentially small. At the same time it is easy to see  that the resulting exponent is  not optimal. For example we  consider three equal-time points $x_i^\mu=(0,\vec{x}_i)$ with all three mutual distances being much larger than the inverse mass gap
\bea
\label{assumption}
m\ell_i\gg 1, \qquad \ell_1=|\vec{x}_2-\vec{x}_3|,\quad  \ell_2=|\vec{x}_3-\vec{x}_1|,\quad  \ell_3=|\vec{x}_1-\vec{x}_2|.
\eea
Without loss of generality throughout this paper we assume 
\bea
\label{order}
\ell_1\geq \ell_2\geq \ell_3.
\eea
We are interested in calculating exponential suppression of
\bea
\label{3pt}
G_{123}(x_1,x_2,x_3)=\langle \phi_1(x_1) \phi_2(x_2) \phi_3(x_3)\rangle.
\eea 
The composite operator $\phi_1(x_1)\phi_2(x_2)$ can be thought of as a sum of local operators, $\phi_1(x_1)\phi_2(x_2)=\sum_k f(x_1,x_2)\phi_k(x_1)$.
In the limit $|x_3-x_1|\gg |x_1-x_2|$ this idea can be made precise with help of the OPE decomposition. We note that while representing $\phi_1(x_1)\phi_2(x_2)$ as a local operator, between two points $x_1,x_2$ we have to choose the point closer to $x_3$. Then in the limit $m\ell_2\gg 1$ the correlation function can be bounded with help of \eqref{2pt}
\bea
\label{3ptnaive}
|\langle \phi_1(x_1) \phi_2(x_2) \phi_3(x_3)\rangle|\leq e^{-m \ell_2},\qquad m\ell_2\gg 1.
\eea
It is easy to see though that the exponential factor in \eqref{3ptnaive} is too naive. Indeed, when all three points are simultaneous $x_i^{0}=0$, without loss of generality we can choose the coordinate system such that point $\vec{x}_3$ sits at the origin, while $\vec{x}_2=(\ell_1,0,0,\dots)$ and $\vec{x}_1=(a,b,0,\dots)$, where 
\bea
a={\ell_1^2 + \ell_2^2 - \ell_3^2\over 2 \ell_1},\qquad b= {\sqrt{\mathcal D}\over 2\ell_1},
\eea
and
\bea
\nonumber
{\mathcal D}&=&2(\ell_1^2\ell_2^2+\ell_2^2\ell_3^2+\ell_3^2\ell_1^2)-\ell_1^4-\ell_2^4-\ell_3^4=\\&& (\ell_1+\ell_2-\ell_3)(\ell_2+\ell_3-\ell_1)(\ell_3+\ell_1-\ell_2)(\ell_1+\ell_2+\ell_3)=16S^2(\ell_1,\ell_2,\ell_3). \label{discr}
\eea
(Here $S(\ell_1,\ell_2,\ell_3)$ is the area of the triangle with the sides $\ell_1,\ell_2,\ell_3$ given by Heron's formula.)
This is shown in Fig.~\ref{fig:triangle}, where we only keep first two components of $\vec{x}_i$, while all others, as well as time component, are identically zero. 
Next, one can use Euclidean quantization and choose time direction along $(x_2-x_3)^\mu$,
\bea
G_{123}=\langle 0| \phi_2(\ell_1,0) \phi_1(a,b) \phi_3(0,0)|0\rangle=\langle 0| \phi_2(0,0) e^{-(\ell_1-a) H} \phi_1(0,b) e^{-a H}\phi_3(0,0)|0\rangle, 
\eea
resulting in the exponential suppression $e^{-m\ell_1}$.  This is better than \eqref{3ptnaive}.
This simple exercise shows that the exponential rate of suppressed of higher-point correlators  imposed by the two-point function is not optimal. In this paper we argue that the optimal rate of suppression, i.e.~the best rate which  would universally apply to all theories and operators $\phi_i$, for the three-point correlator is given by the sum of distances to the operator locations from the Fermat-Steiner point \eqref{Fermatdistance},
\bea
\label{otvet}
\langle \phi_1(x_1) \phi_2(x_2) \phi_3(x_3)\rangle \sim e^{-m\, \ell_{\rm Fermat}}.
\eea
This behavior was previously established in the context of certain two-dimensional models \cite{chim1992integrable,caselle2006potts,delfino2010three}. When the theory is confining it readily follows from the minimal length geometry of flux tubes \cite{takahashi2001three, de2005baryon}, leading to the so-called Y-law. 
We extend \eqref{otvet} to non-confining theories in any dimensions. We also consider configurations when the three points $x^\mu_i$ do not lie on the same spatial plane, while all three mutual intervals are space-like, and introduce the notion of Fermat point in that case. We argue that the suppression rate of the higher point correlation functions is determined by the shortest tree-level graph connecting all points -- the solution of the Euclidean Steiner tree problem.

This paper is organizes as follows. In the next section we discuss three-point function when all three points $x_i^\mu$ belong to the same spatial plane. In section \ref{kinematics} we discuss possible configurations of three points in the Minkowsi space when all mutual intervals are space-like. Section \ref{general} is devoted to calculation  of the suppression rate of the three point function for a general Minkowskian configuration. Section \ref{discussion} concludes with a discussion of higher point functions and implications for operator growth in relativistic theories.

\section{Euclidean configuration}
\label{Euclidean}
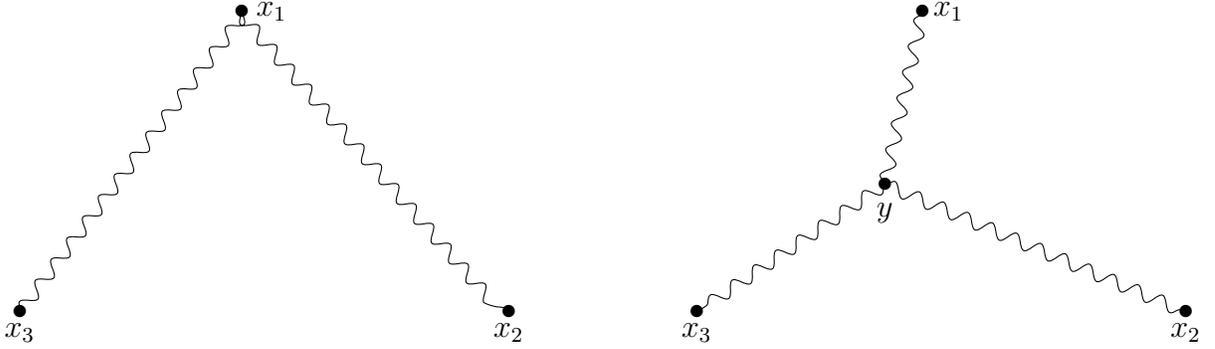
\begin{figure}[t]
\centering
\begin{tikzpicture}
\draw [draw=black,snake it]
    (0,0) node[below] {$x_3$} -- (3,4) node[right] {$x_1$} -- (6.5,0) node[below] {$x_2$};
\node at (0,0) {\textbullet};
\node at (2.95,4) {\textbullet};
\node at (6.5,0) {\textbullet};

\draw [draw=black,snake it]     (11.5,1.7) node [below=3pt] {$y$} -- (9,0) node[below] {$x_3$} ;
\draw [draw=black,snake it]     (11.5,1.7) -- (12,4) node[right] {$x_1$} ;
\draw [draw=black,snake it]    (11.5,1.7) --  (15.5,0) node[below] {$x_2$} ;

\node at (9,0) {\textbullet};
\node at (12.,4) {\textbullet};
\node at (15.5,0) {\textbullet};
\node at (11.5,1.7) {\textbullet};
\end{tikzpicture}
\caption{Simplest Feynman diagrams contributing to the connected part of \eqref{3pt}.}
\label{fig:FD}
\end{figure}

We start with the case when all points belong to a spatial plane, such that all three operators are simultaneous   $x_i^0=0$. We already know that in this case the exponential suppression factor  is not smaller than $\ell_1$. To establish the optimal rate of suppression we consider simplest Feynamn diagrams contributing to the connected part of \eqref{3pt}.
First class of diagrams include no additional vertexes, but only propagators directly connecting some of the operators $\phi_i$. These diagrams are 
present in all theories, including non-interacting ones.  One of these diagrams is schematically depicted in Fig.~\ref{fig:FD} (left). Given that each propagator $G(x_i-x_j)$ is suppressed as $e^{-m|\vec{x}_i-\vec{x}_j|}$, the optimal (largest universal) suppression is given by $e^{-m(\ell_2+\ell_3)}$. This is for example the suppression rate in a theory of free massive scalar field $\varphi$ when $\phi_1=\varphi$, $\phi_2=\varphi^2$, $\phi_3=\varphi$.

Another class of Feynman diagrams include one interaction vertex connected by propagators with the original operators, 
see Fig.~\ref{fig:FD} (right),
\bea
\label{I123}
I_{123}=\int d^d y\, V(-\partial_{x_i})\, G(y-x_1) G(y-x_2) G(y-x_3).
\eea 
Here $V$ is a polynomial in derivatives acting on each ``leg.'' It depends on interaction.  
Since we are only interested in the exponential factor the derivatives can be neglected and $V$ can be substituted by a coupling constant. In principle, the propagators in \eqref{I123} can be different, we only assume that $m$ is the lightest excitation propagating in each channel. A crucial simplification comes from the fact that all operators are simultaneous and the integral \eqref{I123} can be calculated in the Euclidean space. 
Using K\"all\'en-Lehmann representation in the coordinate space 
\bea
\label{EG}
G(x)=\int_m^\infty d\mu^2 \rho(\mu^2)\, \frac{1}{2\pi}\left({\mu\over 2\pi |x|}\right)^{(d-2)/2} K_{(d-2)/2}(\mu |x|),\qquad |x|^2=\sum_\mu x_\mu^2,
\eea
the integral of interest reduces to 
\bea
\label{KKK}
\int d^d y\, {K_{(d-2)/2}(\mu_1 |y-x_1|)K_{(d-2)/2}(\mu_2 |y-x_2|) K_{(d-2)/2}(\mu_3 |y-x_3|)\over (|y-x_1||y-x_2||y-x_3|)^{(d-2)/2}},\quad \mu_i\geq m.\ 
\eea
The integral over $d$-dimensional Euclidean space can be split into the integral over three ball regions $\mu_i|y-x_i|\lesssim 1$ and the rest. Using the assumption $m\ell_i\gg 1$, when $y$ comes close to one of the operators two other propagators can be bounded by exponents, 
\bea
\int\limits_{\mu_1|y-x_1|\lesssim 1} d^d y\, G(y-x_1) G(y-x_2) G(y-x_3)\leq e^{-m(\ell_2+\ell_3)} \int\limits _{\mu_1|y-x_1|\lesssim 1} d^d y\, G(y-x_1), \ 
\eea
where we neglected order one factors which do not affect the leading exponent and also have used that for $t\gg 1$,
\bea
\label{Kineq}
0< K_{(d-2)/2}(t)< e^{-t}.
\eea
The integral 
\bea
\int\limits_{m|y-x_1|\lesssim 1} d^d y\, G(y-x_1) \label{1pt}
\eea 
may diverge in the UV  if the UV dimension of $\phi_1$ is equal or greater than $d/2$. In \eqref{KKK} this divergence will manifest after integrating over $\mu_1$. In this case the integral in \eqref{1pt} has to be regularized by introducing an appropriate UV-cutoff. Importantly, \eqref{1pt} does not depend on other points $x_2,x_3$ and upon regularization will become some $\ell_i$-independent constant. Thus, we conclude that the integral \eqref{I123} over the ball regions around the original operators will give the same exponential suppression factor $e^{-m(\ell_2+\ell_3)}$ as the ``non-interacting'' Feynman diagrams discussed above and shown  in Fig.~\ref{fig:FD} (left).

The integral \eqref{KKK} over the rest of the $d$-dimensional space excluding the balls $\mu_i|y-x_i|\lesssim 1$ can be bounded by 
\bea
\int d^d y\, e^{-m(|y-x_1|+|y-x_2|+|y-x_3|)}, \label{exponent}
\eea
where we used \eqref{Kineq} and $\mu_i\geq m$. This integral can be extended back to the whole Euclidean $d$-dimensional space, because the additional ``added by hands'' integrals of the exponent  $e^{-m(|y-x_1|+|y-x_2|+|y-x_3|)}$ over the regions $m|y-x_i|\lesssim 1$ is suppressed by $e^{-m(\ell_k+\ell_l)}$, $i\neq k,l$ and thus unimportant.   The leading (optimal) exponent is given by the smallest value 
\bea
\label{minimum}
\min_{y\in {\mathbb R}^d}\, |y-x_1|+|y-x_2|+|y-x_3|. 
\eea
Clearly the minimum is achieved when $y$ belongs to the same two-dimensional spatial plane as $x_i$. Hence minimization problem \eqref{minimum} becomes the famous Fermat-Torricelli problem of finding a point on a plane such that total distance from the three vertexes of a given triangle to that point is the minimum possible. It is easy to see that the minimal total distance, which we denote $\ell_{\rm Fermat}$ is not larger than $\ell_2+\ell_3$. Hence all terms suppressed as $e^{-m (\ell_2+\ell_3)}$ are subleading, while the optimal exponent is given by \eqref{otvet}. The expression for $\ell_{\rm Fermat}$ in terms of $\ell_i$ will be given in \eqref{Fermatdistance} below. 

The derivation above relied on the fact that all three points belong to a spatial plane, hence the integral \eqref{I123} can be written in the Euclidean space. This is not always possible, even if all mutual intervals are space-like. In the next sections we consider the general case and extend the notion of the Fermat point when the corresponding triangle is Minkowskian. 

So far we have only considered the simples Feynman diagrams which corresponds to the first order of perturbation theory.  In fact this is sufficient to establish the result non-perturbatively. We first discuss the case of the three-point function of fundamental fields below. The case of composite operators is discussed in the Appendix A. 

The correlator of fundamental fields $\langle \varphi_1(x_1) \varphi_2(x_2) \varphi_3(x_3)\rangle$ can be calculated exactl using the effective action formalism \cite{goldstone1962j,weinberg1995quantum2}. Then the full connected correlator is given by  the tree-level diagram shown in the right panel of Fig.~\ref{fig:FD}, which can be written as \eqref{I123}, with  $G_\varphi$ being the dressed propagator \eqref{EG} and the interaction vertex $V$ being the sum of all 1PI diagrams with three legs. In the regime $m|x_i-y|\gtrsim 1$ the propagator can be approximated by $e^{-m|x_i-y|}$, hence the effective value of the corresponding momentum $p_i\sim\partial_i$ is of order $m$.  In other words the calculation in the effective theory reduces to the calculation of the tree-level diagram discussed above with the effective value of the coupling constant $V(p_i\approx m)$. The latter can be large if the theory is strongly coupled in the IR. Importantly, this large factor is not $\ell_i$-dependent, i.e.~it remains the same while the mutual distances between three points $x_i^\mu$ are taken to infinity. Thus we find 
in full generality the asymptotic behavior to be
\bea
|\langle \varphi_1(x_1) \varphi_2(x_2) \varphi_3(x_3)\rangle| \lesssim V(m) e^{-m\, \ell_{\rm Fermat}},\qquad m\ell_i\gg 1.
\eea



\section{Kinematics of three points in the Minkowski space}
\label{kinematics}
Our goal in this section is to consider all possible configurations of three points $x_i^\mu$ in the Minkowski space with the signature $(+,-,-,\dots)$, assuming their mutual intervals are space-like, 
\bea
\label{distances}
(x_2-x_3)^2=-\ell_1^2,\quad (x_3-x_1)^2=-\ell_2^2,\quad  (x_1-x_2)^2=-\ell_3^2,\qquad 
\ell_1\ge \ell_2\geq \ell_3.
\eea
Three points always belong to a two-dimensional plane spanned by the vectors 
\bea
u^\mu=x^\mu_2-x^\mu_3,\qquad v^\mu=x^\mu_1-x^\mu_3.
\eea
The signature of the embedded metric is given by the Gram matrix 
\bea
g=\left(
\begin{array}{cc}
u^\mu u_\mu & u^\mu v_\mu \\
u^\mu v_\mu & v^\mu v_\mu 
\end{array}
\right)=
\left(
\begin{array}{cc}
-\ell_1^2 & (\ell_3^2-\ell_1^2-\ell_2^2)/2 \\
(\ell_3^2-\ell_1^2-\ell_2^2)/2 & -\ell_2^2
\end{array}
\right).
\eea
The trace of Gram matrix is negative, which means that at least one of the directions is space-like. The signature of the other direction follows from the determinant,
\bea
\det g=-{\mathcal D\over 4}.
\eea 
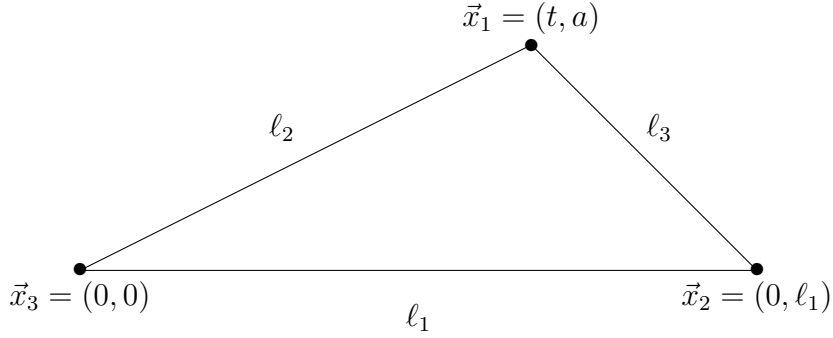
\begin{figure}[t]
\centering
\begin{tikzpicture}
\draw
    (0,0) coordinate (a) node[below] {$\vec{x}_3=(0,0)$}
    -- (9,0) coordinate (b) node[below] {$\vec{x}_2=(0,\ell_1)$}
    -- (6,3) coordinate (c) node[above] {$\vec{x}_1=(t,a)$}
    -- (a);
\draw (4.5,-0.3) coordinate () node[below] {$\ell_1$};
\draw (7.7,1.9) coordinate () node {$\ell_3$};
\draw (3,1.9) coordinate () node[left] {$\ell_2$};
\node at (0,0) {\textbullet};
\node at (6,2.98) {\textbullet};
\node at (9,0) {\textbullet};
\end{tikzpicture}
\caption{Simplest kinematics when the triangle inequality is violated $\ell_2+\ell_3<\ell_1$ and three points are inherently Minkowskian.}
\label{fig:M3}
\end{figure}
As follows from \eqref{discr}, the corresponding plane is space-like iff $\ell_1,\ell_2,\ell_3$ satisfy all three triangle inequalities 
$\ell_1< \ell_2+\ell_3,\ \ell_2< \ell_3+\ell_1,\ \ell_3< \ell_1+\ell_2$. If we assume without loss of generality that  $\ell_i$ are ordered as in \eqref{order}, then the points belong to a spatial plane whenever $\ell_2+\ell_3>\ell_1$. In this case time coordinate of all three points can be chosen to be zero, while other coordinates can be brought to the form as in Fig.~\ref{fig:triangle}. If the triangle inequality is not satisfied $\ell_2+\ell_3<\ell_1$, it is impossible to choose the coordinate system such that all three points are simultaneous. In this case the simplest kinematics is achieved in the coordinate system such that $x_3^\mu$ sits at the origin, $x_2^\mu=(0,\ell_1,0,\dots)$, and $x_1^\mu=(t,a,0,\dots)$, where 
\bea
a={\ell_1^2 + \ell_2^2 - \ell_3^2\over 2 \ell_1},\qquad t= {\sqrt{- \mathcal D}\over 2\ell_1},
\eea
see Figure~\ref{fig:M3}.

An interesting situation is when $\ell_2+\ell_3=\ell_1$. In this case one of the directions is light-like, and the general configuration can be brought to the form 
\bea
\label{lightlike}
x_1^\mu=(t,\ell_2,t,\dots),\qquad x_2^\mu=(0,\ell_1,0,\dots),\qquad  x_3^\mu=(0,0,0,\dots),
\eea
where the parameter $t$ could be either zero or can be brought to be $t=1$.

\section{General configuration}
\label{general}
\begin{figure}
\centering
\begin{tikzpicture}
\draw
    (0,0) coordinate (a) node[below] {$\vec{x}_3=(0,0)$}
    -- (9,0) coordinate (b) node[below] {$\vec{x}_2=(0,\ell_1)$}
    -- (6,4) coordinate (c) node[above] {$\vec{x}_1=(a,b)$}
    -- (a);
\draw (4.5,-0.3) coordinate () node[below] {$\ell_1$};
\draw (7.7,2.3) coordinate () node {$\ell_3$};
\draw (3,2.3) coordinate () node[left] {$\ell_2$};
\node at (0,0) {\textbullet};
\node at (6,3.98) {\textbullet};
\node at (9,0) {\textbullet};
\node (F) at (5.8,2.43) {\textbullet};
\draw[dashed]
    (a) --(F)--(b); 
\draw[dashed]
    (F)--(c);
\draw (3.5,1.3) coordinate () node[below] {$2m\gamma$};
\draw (7.,1.3) coordinate () node[below] {$2m\beta$};
\draw (5.45,3.5) coordinate () node[below] {$2m\alpha$};

\draw (F) circle (0.38cm);
\draw (F) circle (0.42cm);

\draw (5.8,2.1) coordinate () node[below] {\small $120\degree$};
\draw (5.1,3) coordinate () node[below] {\small $120\degree$};
\draw (6.6,2.8) coordinate () node[below] {\small $120\degree$};

\end{tikzpicture}
\caption{The sides $x_i-x_j$ of the original triangle  together with the lines connecting Fermat point  with the original points $x_i$ form three triangles, each has an obtuse angle of $120\degree$.}
\label{fig:Fermat}
\end{figure}
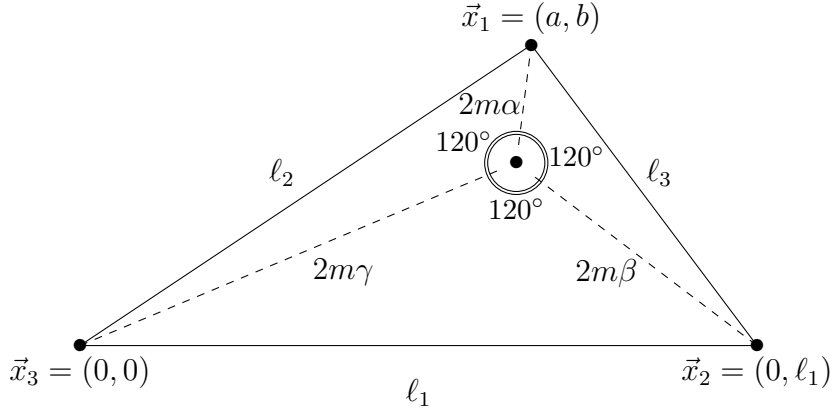

To estimate the leading exponent in case when the configuration is Minkowskian we resort to a massive $\varphi^3$ theory when all three operators are the same $\phi_i=\varphi$. Then the integral \eqref{I123} written in the momentum space is given by 
\bea
I_{123}=\int {d^d k_1 \over (2\pi)^d} \int  {d^d k_2 \over (2\pi)^d} {e^{ik_1(x_1-x_3)+ik_2(x_2-x_3)}\over (k_1^2-m^2+i\epsilon)(k_2^2-m^2+i\epsilon)((k_1+k_2)^2-m^2+i\epsilon)}.
\eea
Using Schwinger parameter representation we can reduce the integral to 
\bea
\label{i123}
I_{123}=\frac{i^{d-3}}{(4 \pi)^d}\int_0^\infty d\alpha \int_0^\infty d\beta \int_0^\infty d\gamma \frac{1}{(\alpha \beta+\beta\gamma+\gamma\alpha)^{d/2}}\nonumber\\ {\rm exp}\left(-{i\over 4}{\alpha \ell_1^2 +\beta\ell_2^2+\gamma \ell_3^2\over \alpha \beta+\beta\gamma+\gamma\alpha}+i(\alpha+\beta+\gamma)(-m^2+i \epsilon)\right).
\eea
The main contribution comes from the saddle point, 
\bea
\label{geometry1}
\ell_1^2  = 4 m^2 (\beta^2 + \gamma^2 + \beta \gamma),  \\
\ell_2^2 = 4 m^2 (\gamma^2 + \alpha^2 + \gamma \alpha ),  \\
\ell_3^2= 4 m^2  (\alpha^2 + \beta^2 + \alpha \beta),  \\
\label{geometry3}
\eea
provided it belongs to the octant $\alpha,\beta,\gamma\geq 0$. The equations (\ref{geometry1}-\ref{geometry3}) have a simple geometric interpretation. It is a law of cosine for the triangles which has one angle equal 120$^\circ$ and the largest side (opposite to that angle) being one of the $\ell_i$, while two other sides being $2m$ multiplied by 
 $\alpha$, $\beta$, or $\gamma$. In other words, $2m$ multiplied by $\alpha, \beta, \gamma$ give the distances from the original points $x_i$ to the point from which each side  of the corresponding triangle is ``seen'' at 120$^\circ$. Such a point exists only for triangles where the largest angle is less than 120$^\circ$ and, when it exists, it is the Fermat point. This is shown in Fig.~\ref{fig:Fermat}.

To further simplify \eqref{i123} we introduce Feynman parameters $a,b,c$ subject to constraints 
\bea
\label{triangle}
a,b,c\geq 0, \qquad a+b+c=1,
\eea
and Schwinger parameter $t$,
\bea
\alpha=t a,\qquad  \beta=t b, \qquad  \gamma=t c.
\eea
The integral over $t$ can be calculated, yielding 
\bea
\label{xyzint}
I_{123}=\frac{2(i m)^{d-3}}{(4 \pi)^d}\int_{\Delta} da\, db\, \frac{ K_{d-3}(m l)}{l^{d-3}(a b +b c+ ca)^{d/2}},\qquad l^2={a\, \ell_1^2+b\, \ell_2^2+c\,\ell_3^2\over ab +bc +ca}.
\eea
The integral in \eqref{xyzint} is over the triangle \eqref{triangle}. Macdonald function  $K_{d-3}(t)$ is positive definite and can be approximated by an exponent for large values of the argument. It is thus clear that the integral \eqref{xyzint}  in the limit 
$m\ell_i\gg 1$ is saturated by the maximal value of $l(a,b,c)$ inside the triangle \eqref{triangle}. 
We start our analysis with the conventional Euclidean case when $\ell_2+\ell_3>\ell_1$.
When all angles of the original triangle are smaller than $120\degree$, i.e. $\ell_2^2+\ell_2\ell_3+\ell_3^2>\ell_1^2$, the Fermat point is located strictly inside the original triangle, all three lengths $\alpha,\beta,\gamma>0$ and $l(a,b,c)$ achieves it maximum  inside \eqref{triangle},
\bea
\max_{\Delta}\, l(x,y,z)=\ell_{\rm Fermat}= \sqrt{\frac{1}{2}(\ell^2_1+\ell_2^2+\ell_3^2 + \sqrt{3 \mathcal D})}.
\eea
When $(\ell_2^2+\ell_2\ell_3+\ell_3^2)/\ell_1^2$ decreases and becomes smaller than $1$,  the obtuse angle becomes equal or large $120\degree$, then the Fermat point coincides with the vertex of the obtuse angle $x^\mu_1$. In this case the maximum of $l(a,b,c)$ is achieved at the boundary $a=0$,
\bea
\label{boundary}
\max_{\Delta}\, l(a,b,c)=\ell_{\rm Fermat}=\ell_2+\ell_3,
\eea
and the contributions of both Feynman diagrams depicted in Fig.~\ref{fig:FD} is of the same order. When $\ell_1$ further grows and approaches $\ell_1=\ell_2+\ell_3$ the triangle degenerates into a spatial line or belongs to a plane with one direction being light-like, see \eqref{lightlike}. When $\ell_1>\ell_2+\ell_3$ the triangle inequality is violated and the triangle is inherently Minkowskian, see Fig.~\ref{fig:M3}. In all cases $\ell_1\geq \ell_2+\ell_3$ the maximum of $l(a,b,c)$ is achieved on the boundary $a=0$ and is given by \eqref{boundary}. In other words we can define Fermat point for all cases when  $(\ell_2^2+\ell_2\ell_3+\ell_3^2)\leq \ell_1^2$ as being the vertex of the ``obtuse angle'' $x_1^\mu$. Finally we have 
\bea
\ell_{Fermat}=\left\{ 
\begin{array}{cr}
\sqrt{\frac{1}{2}(\ell^2_1+\ell_2^2+\ell_3^2 + \sqrt{3 \mathcal D})}, & \quad \ell_2^2+\ell_2\ell_3+\ell_3^2> \ell_1^2, \\
\ell_2+\ell_3, & \ell_2^2+\ell_2\ell_3+\ell_3^2\leq \ell_1^2.
\end{array}
\right.
\label{Fermatdistance}
\eea

\section{Discussion}
\label{discussion}
In this paper, we have argued that the connected part of a three-point function in a general relativistic quantum field theory with a mass gap $m$ decays as 
\bea
\label{answer}
\langle \phi_1(x_1) \phi_2(x_2) \phi_3(x_3)\rangle \sim e^{-m\, \ell_{\rm Fermat}},
\eea
where $\ell_{\rm Fermat}$ \eqref{Fermatdistance} is the total distance from the Fermat point to the operator locations $x_i$. When the mutual distances \eqref{distances} satisfy $(\ell_2^2+\ell_2\ell_3+\ell_3^2)\leq \ell_1^2$ (this also includes  all inherently Minkowksian configurations), the Fermat point coincides with the edge of the obtuse angle $x_1$. We first considered purely Euclidean configurations and established \eqref{answer} at first order in perturbation theory, while also explaining why 
exactly the same calculation  is valid non-pertrubatively for to the calculation in the effective theory.
Then we calculated the corresponding Feynman diagram explicitly for an arbitrary configuration of $x_i^\mu$ in the theory of massive scalar field and saw that continuation into Minkowski space does not dramatically change the result. 

It is an interesting question to extend our consideration to a general $n$-point function $\langle \phi_1(x_1)\dots \phi_n(x_n)\rangle$ assuming all mutual intervals $x_i-x_j$ are space-like. When all operators are simultaneous, $x_i^0=0$, the problem can be formulated in the Euclidean space, and similarly to the discussion in section \ref{Euclidean}, we expect that the largest (least suppressed) term will be given by a tree-level Feynman diagram connecting all original points and the additional vertexes, such that the total Euclidean distance of the corresponding graph would me minimal. In other words, the leading exponent will be given by the total length of the graph solving the Euclidean Steiner problem \cite{jarnik1934minimalnich}. Such a graph typically consists of the Fermat-Steiner points each connecting three lines, with all angles being 120\degree.  This similarity with the three-point case suggests that for the Minkowsian configurations, when the choice $x_i^0=0$ is not possible, the answer will only change in the way that some of the graph vertexes will merge. It should be noted, that finding optimal graph of the Euclidean Steiner problem, and thus in turn calculating higher order correlation function, is NP hard \cite{bern1989shortest,brazil2014history}. It would be interesting to understand the interplay between the complexity of  corresponding optimization problem and elegance of the close cousins of correlation functions -- tree-level scattering amplitudes in field theory \cite{Britto:2005fq,ArkaniHamed:2008yf,Arkani-Hamed:2013jha,Arkani-Hamed:2017jhn}. It should be also noted that the Steiner problem in the hyperbolic space appears in the holographic context as an effective description of large $c$ conformal blocks
\cite{alkalaev2018perturbative}.
Another curious connection is between optimal Euclidean trees and minimal surfaces (soap films) \cite{isenberg1975problem}. We leave it as an intriguing question for the future to  explore if these connections may lead to new computational techniques or optimization algorithms.

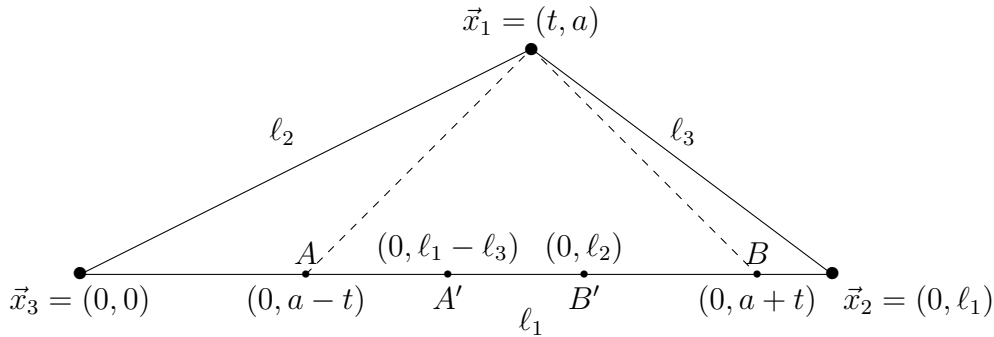
\begin{figure}[t]
\centering
\begin{tikzpicture}
\draw
    (0,0) coordinate (a) node[below] {$\vec{x}_3=(0,0)$}
    -- (10,0) coordinate (b) node[below right] {$\vec{x}_2=(0,\ell_1)$}
    -- (6,3) coordinate (c) node[above] {$\vec{x}_1=(t,a)$}
    -- (a);
\draw [dashed] (3,0) node[below] {$(0,a-t)$} -- (6,3)--(9,0) node[below] {$(0,a+t)$};

\node[below] at (4.89,0) {$A'$};
\node[above] at (4.89,0) {$(0,\ell_1-\ell_3)$};
\node at (4.89,0) {\tiny \textbullet};
\node[below] at (6.7,0) {$B'$};
\node[above] at (6.7,0) {$(0,\ell_2)$};
\node at (6.7,0) {\tiny \textbullet};

\node[above] at (3,0) {$A$};
\node at (3,0) {\tiny \textbullet};
\node[above] at (9,0) {$B$};
\node at (9,0) {\tiny \textbullet};

\draw (6,-0.3) coordinate () node[below] {$\ell_1$};
\draw (7.7,1.9) coordinate () node[right] {$\ell_3$};
\draw (3,1.9) coordinate () node[left] {$\ell_2$};
\node at (0,0) {\textbullet};
\node at (6,2.98) {\textbullet};
\node at (10,0) {\textbullet};
\end{tikzpicture}
\caption{Time evolved operator $\phi_1(t,a)=e^{-iHt}\phi_1(0,a) e^{iHt}$ at time $t=0$ is confined to the light-cone area inside the $AB$ interval.}
\label{fig:spreading}
\end{figure}

Many-point correlation functions considered in this paper, with all operators mutually commuting, can be understood as a very simple cousins of the out of time ordered correlation functions \cite{larkin1969quasiclassical,shenker2014black,maldacena2016bound}, which are efficient probes of the operator growth and scrambling \cite{roberts2016lieb,von2018operator,nahum2018operator,khemani2018operator,roberts2018operator,Qi}. 
Similarly, our results also have interesting implications for the operator growth in relativistic theories.
We start with the two point function. 
Consider two operators located at  the origin $x_2^\mu=(0,0)$ and $x_1^\mu=(t,a)$, $t<a$, (for simplicity we consider two-dimensional Minkowski space). The operator $\phi_1(t,a)=e^{-iHt}\phi_1(0,a)e^{itH}$ can be understood as a non-local operator at time $t=0$ spread inside the light-cone region from $a-t$ to $a+t$. The detailed structure of this operator is complicated, but {inside the correlator $\langle \phi_1\phi_2\rangle$}, from the point of view of the operator $\phi_2(0,0)$ operator $\phi_1(t,a)$ will be perceived as a local operator sitting at $(0, \sqrt{a^2-t^2})$. This is an exact result dictated by Lorentz symmetry. Now we consider three operators inserted at the points $x_i^\mu$ shown in Figure \ref{fig:spreading}. The operator $\phi_1(t,a)=e^{-iHt}\phi_1(0,a)e^{itH}$ is spread between the points $A=(0,a-t)$ and $B=(0,a+t)$ at time $t=0$ and is non-local.  The leading behavior of the 3-point correlation function is given by  \eqref{boundary}, 
\bea
\langle \phi_1(t,a)\phi_2(0,\ell_1)\phi_3(0,0)\rangle \sim e^{-m(\ell_2+\ell_3)}.
\eea
Hence, at leading order, $\phi_1(t,a)$ is perceived by $\phi_2(0,\ell_1)$ and $\phi_3(0,0)$ as a local operator sitting at the points  $A'=(0,\ell_1-\ell_3)$ and  $B'=(0,\ell_2)$ correspondingly. These points are inside the lightcone of $x_1=(t,a)$ bounded by points $A$ and $B$ and satisfy the condition that the distance between $x_2$ and $A'$ plus the distance between $x_3$ and $B'$ is equal to $\ell_2+\ell_3$.
This result would not be surprising (and would be exact) if we considered the disconnected contributions associated with $\langle \phi_1 \phi_2\rangle \langle \phi_3\rangle $ and  $\langle \phi_1 \phi_3\rangle \langle \phi_2\rangle $. Rather, this applies to the connected part of  $\langle \phi_1 \phi_2 \phi_3\rangle $ and is no longer guaranteed by Poincare symmetry. Perceived locality of $e^{-iHt}\phi_1(0,a)e^{itH}$ suggests a particular structure of the time-evolved local operators in relativistic theories. 

An important and interesting question would be to generalize our results to lattice systems. While the exponential rate of cluster decomposition for two operators is understood \cite{hastings2004lieb,hastings2006solving,hastings2010locality}, generalization to three and more operators is a non-trivial task. Borrowing from the technique of establishing correlation length at finite temperatures \cite{kliesch2014locality,huang2018lieb}, we expect the leading exponent to be given by the shortest path on the lattice connecting all operators, i.e.~rectilinear Steiner tree problem for cubic lattices. Furthermore, operator growth in lattice models with short-range interactions exhibit an emergent light-cone structure \cite{lieb1972finite,kim2013ballistic,luitz2017information,von2018operator,nahum2018operator,khemani2018operator} and in many models full relativistic symmetry is known to emerge at large distances. It would be very interesting to understand the microscopic origin of the perceived locality of $e^{-iHt}\phi_1(0,a)e^{itH}$ from the point of view of the other operators in the correlation function in such models.

\acknowledgments
We thank Slava Rychkov and David Simmons-Duffin for discussions. 
This work is supported by the BSF grant 2016186. 
The work of LA is supported by the Russian Academic Excellence Project `5-100'.
AD is grateful to KITP for hospitality, where this work has been completed. 
The research at KITP was supported in part by the National Science Foundation under Grant No. NSF PHY-1748958.

\appendix
\section{Composite operators}

In this Appendix we consider a three-point function of renormalized composite operators  $\langle [\mathcal{O}_1] (x_1) ~ [\mathcal{O}_2] (x_2) ~ [\mathcal{O}_3] (x_3)\rangle$. As usual, the renormalized $[\mathcal{O}_i](x)$ is given by the mixing of bare $\mathcal{O}_i(x)$ with all possible operators of lower dimension. In the minimal subtraction scheme, the coefficients of various operators participating in the definition of $[\mathcal{O}_i](x)$ contain an ascending series of poles such that insertion of the renormalized operator into any correlation function of the fundamental fields is finite. 

Our analysis is parallel to the discussion of $\langle \varphi(x_1) \varphi(x_2) \varphi(x_3)\rangle$ in the main body of the text with one important exception. While in the case of fundamental fields it is enough to focus on the three-point interaction vertex of the effective action, in the case of composite operators it is necessary to account for all possible skeleton diagrams with external legs matching fundamental fields in the composite operators $\mathcal{O}_i(x),~ 1\leq i\leq3$. 

\begin{figure}
\begin{subfigure}{0.45\textwidth}
\begin{tikzpicture}
\draw [draw=black,snake it]
    (0,0) node[below] {$x_3$} -- (3,4) node[right] {$x_1$} -- (6.5,0) node[below] {$x_2$} -- (0,0);
\node at (0,0) {\textbullet};
\node at (2.95,4) {\textbullet};
\node at (6.5,0) {\textbullet};
\end{tikzpicture}
\caption{Tree-level diagram in the effective theory without interacting vertexes.}
\label{fig:tree}
\end{subfigure}
\begin{subfigure}{0.05\textwidth}
\end{subfigure}
\begin{subfigure}{0.5\textwidth}
\begin{tikzpicture}
\node[circle,text width=2cm,fill=gray!20] (6v) at (6,0) {\centering $6$-point \\ \quad vertex };
\node[] (left) at (3,0) {$p_1$};
\node[] (right) at (9,0) {$q_2$};
\node[] (rightup) at (8.,2.4) {$p_2$};
\node[] (leftup) at (4.,2.4) {$q_1$};
\node[] (rightdown) at (8.,-2.4) {$p_3$};
\node[] (leftdown) at (4.,-2.4) {$q_3$};
\draw[->,thick,draw=black!50,line width=1mm] (right.west) to [out=180,in=0] (6v.east);
\draw[->,thick,draw=black!50,line width=1mm] (left.east) to [out=0,in=180] (6v.west);
\draw[->,thick,draw=black!50,line width=1mm] (rightup.south west) to [out=240,in=60] (6v.north  east);
\draw[->,thick,draw=black!50,line width=1mm] (leftup.south east) to [out=300,in=120] (6v.north  west);
\draw[->,thick,draw=black!50,line width=1mm] (rightdown.north west) to [out=120,in=300] (6v.south  east);
\draw[->,thick,draw=black!50,line width=1mm] (leftdown.north east) to [out=60,in=240] (6v.south  west);
\node () at (6,-2) {$V_6$};
\end{tikzpicture}
\caption{$V_6$ vertex represents a sum over all 1PI Feynman diagrams with six external legs. The momenta are taken to flow inwards.}
\label{fig:V6}
\end{subfigure}
\end{figure}
\begin{figure}
\begin{tikzpicture}
\node[] () at (0,0) {};
\node[circle,text width=1cm,fill=gray!20] (3c) at (3,0.4) {$\, \, \, \, V_3$};
\node[circle,text width=1cm,fill=gray!20] (3b) at (3,-2) {$\, \, \, \, V_3$};
\node[circle,text width=1cm,fill=gray!20] (3l) at (1.3,2.5) {$ \, \, \, \, V_3$};
\node[circle,text width=1cm,fill=gray!20] (3r) at (4.7,2.5) {$ \, \, \, \, V_3$};
\draw[->,thick,draw=black!50,line width=1mm] (3b.north) to [out=90,in=270] (3c.south);
\draw[->,thick,draw=black!50,line width=1mm] (3l.south east) to [out=300,in=120] (3c.north west);
\draw[->,thick,draw=black!50,line width=1mm] (3r.south west) to [out=240,in=60] (3c.north east);
\node[] () at (4,-1.0) {$p_3+q_3$};
\node[] () at (1.5,1.1) {$p_1+q_1$};
\node[] () at (4.6,1.1) {$p_2+q_2$};
\node[] (rightdown) at (4.,-3.4) {$p_3$};
\node[] (leftdown) at (2.,-3.4) {$q_3$};
\draw[->,thick,draw=black!50,line width=1mm] (2.2,-3.3) to [out=60,in=240] (2.7,-2.55);
\draw[->,thick,draw=black!50,line width=1mm] (3.75,-3.3) to [out=120,in=300] (3.3,-2.55);
\node[] (p1) at (-.5,2.5) {$p_1$};
\node[] (q1) at (1.3,4.2) {$q_1$};
\draw[->,thick,draw=black!50,line width=1mm] (q1.south) to [out=270,in=90] (3l.north);
\draw[->,thick,draw=black!50,line width=1mm] (p1.east) to [out=0,in=180] (3l.west);
\node[] (p2) at (6.5,2.5) {$p_2$};
\node[] (q2) at (4.7,4.2) {$q_2$};
\draw[->,thick,draw=black!50,line width=1mm] (q2.south) to [out=270,in=90] (3r.north);
\draw[->,thick,draw=black!50,line width=1mm] (p2.west) to [out=180,in=0] (3r.east);
\node[circle,text width=1.3cm,fill=gray!20] (33c) at (11,0.2) {$\, \, \, \,  \,  \, V_4$};
\node[circle,text width=1cm,fill=gray!20] (33l) at (9.3,2.5) {$\, \, \, \, V_3$};
\node[circle,text width=1cm,fill=gray!20] (33r) at (12.7,2.5) {$\, \, \, \, V_3$};
\draw[->,thick,draw=black!50,line width=1mm] (33l.south east) to [out=300,in=120] (33c.north west);
\draw[->,thick,draw=black!50,line width=1mm] (33r.south west) to [out=240,in=60] (33c.north east);
\node[] () at (9.5,1.1) {$p_1+q_1$};
\node[] () at (12.6,1.1) {$p_2+q_2$};
\node[] (pp3) at (12.,-2) {$p_3$};
\node[] (qq3) at (10.,-2) {$q_3$};
\draw[->,thick,draw=black!50,line width=1mm] (10.1,-1.6) to [out=60,in=240] (10.7,-0.6);
\draw[->,thick,draw=black!50,line width=1mm] (11.85,-1.6) to [out=120,in=300] (11.3,-.6);
\node[] (pp1) at (7.5,2.5) {$p_1$};
\node[] (qq1) at (9.3,4.2) {$q_1$};
\draw[->,thick,draw=black!50,line width=1mm] (qq1.south) to [out=270,in=90] (33l.north);
\draw[->,thick,draw=black!50,line width=1mm] (pp1.east) to [out=0,in=180] (33l.west);
\node[] (pp2) at (14.5,2.5) {$p_2$};
\node[] (qq2) at (12.7,4.2) {$q_2$};
\draw[->,thick,draw=black!50,line width=1mm] (qq2.south) to [out=270,in=90] (33r.north);
\draw[->,thick,draw=black!50,line width=1mm] (pp2.west) to [out=180,in=0] (33r.east);
\end{tikzpicture}
\caption{Two skeleton diagrams built out of the effective action vertices $V_3$ and $V_4$ with 3 and 4 external legs respectively.}
\label{fig:comp}
\end{figure}
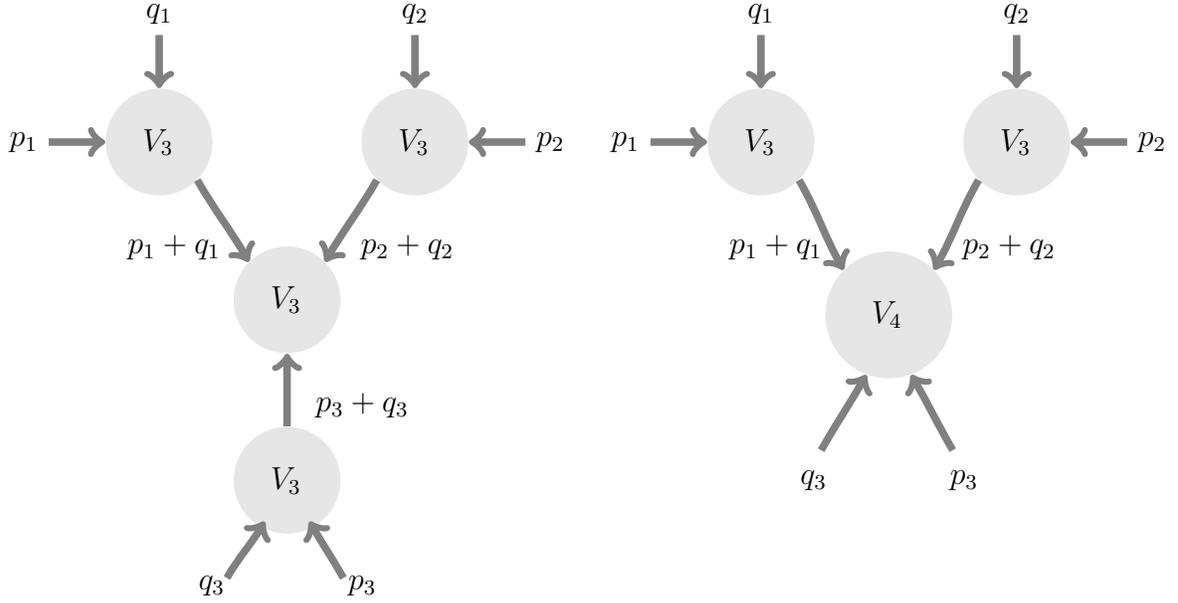

Below we illustrates this idea for a particular choice $\mathcal{O}_i=\varphi^2$.
There are several types of contributions in this case. The simplest skeleton diagram shown in Fig.~\ref{fig:tree} includes no interaction vertexes, while the legs are fully dressed propagators $G_\varphi$ \eqref{EG}. This type of diagrams was discussed in the main text. It is easy to see that it contributes at the subleading order as $e^{-m(\ell_1+\ell_2+\ell_3)}$. Another skeleton diagram includes the interaction vertex given by the sum of all 1PI diagrams with six external legs (sextic vertex in the effective action shown in Fig~\ref{fig:V6}, see \cite{goldstone1962j,weinberg1995quantum2}). Yet other diagrams originate from the connected tree-level diagrams shown in Fig~\ref{fig:comp} (these are skeleton diagrams built from the cubic and quartic vertices of the effective action). Their sum corresponds to an effective sextic vertex which is essential for evaluating the $\langle [\varphi^2](x_1)  ~  [\varphi^2](x_2)   ~  [\varphi^2](x_3)  \rangle$ correlator. In the momentum space the effective vertex takes the form\footnote{Bar over the vertices indicates that we strip off the momentum conservation delta function.}
\begin{eqnarray}
\label{V6}
 &&\bar V^\text{eff}_6(p_i, q_i)= -\bar V_6(p_i, q_i) + 27 \, \bar V_3(p_1+q_1,p_2+q_2,p_3+q_3)
 \prod_{i=1}^3 \bar V_3(p_i,q_i,-p_i-q_i) G_\varphi(p_i+q_i) 
 \nonumber\\
 &&-54\, \bar V_4(p_1+q_1,p_2+q_2,p_3, q_3) \prod_{i=1}^2 \bar V_3(p_i,q_i,-p_i-q_i) G_\varphi(p_i+q_i) ~,
\end{eqnarray}
where in the RHS \eqref{V6} we assume full symmetrization with respect to $p_i,q_i$ and $G_\varphi$ is the dressed propagator \eqref{EG}, whereas $V_3$ and $V_4$ represent the sum of all 1PI diagrams with three and four external legs respectively.

In the regime of large separations $m|x_i-x_j|\gg 1,\, ~ i\neq j$, the structure of the interaction vertices $V_3$, $V_4$ and $V^\text{eff}_6$ tremendously simplifies. As discussed in section \ref{Euclidean}, in this limit the external momenta modes $p_i$ and $q_i$ can be identified with the mass scale $m$ of the excitations created by $\varphi$. In particular,  the full connected correlator in this limit takes the form 
\begin{eqnarray}
&&\langle [\varphi^2] ~ [\varphi]^2 ~ [\varphi_3]^2\rangle \underset{m|x_{i}-x_j|\gg 1}{\longrightarrow} 
6! \, \bar V^\text{eff}_6 (p_i, q_i \approx m) \int d^d y\, G_{\varphi^2}(y-x_1) 
G_{\varphi^2}(y-x_2) G_{\varphi^2}(y-x_3) ~,
\nonumber\\
\end{eqnarray}
where $G_{\varphi^2}(x)$ this time denotes square of the K\"all\'en-Lehmann representation \eqref{EG}.
In other words the calculation reduces to the tree-level diagram discussed in the text with certain $\ell_i$-independent effective coupling constant $V(p_i\approx m)$ and slightly modified propagator. Similar simplification holds for other composite operators and effective vertices. Hence, we find 
in full generality the asymptotic behavior 
\bea
|\langle [\mathcal{O}_1] (x_1) ~ [\mathcal{O}_2] (x_2) ~ [\mathcal{O}_3] (x_3)\rangle| \lesssim V(m)\, e^{-m\, \ell_{\rm Fermat}},\qquad m\,\ell_i\gg 1.
\eea

\bibliographystyle{JHEP}
\bibliography{Fermat}

\end{document}